\documentclass[letterpaper, 10 pt, conference]{ieeeconf}  

\IEEEoverridecommandlockouts                              

\usepackage{microtype}
\usepackage{graphicx}
\usepackage{subfigure}
\usepackage{booktabs} 

\usepackage{epsfig}
\usepackage{booktabs}
\usepackage{multicol}
\usepackage{newfloat}
\usepackage{listings}
\usepackage{multirow}

\usepackage{mathtools}
\usepackage{multirow}
\usepackage[capitalize,noabbrev]{cleveref}

\usepackage[textsize=tiny]{todonotes}

\title{\LARGE \bf
AudioScene: Integrating Object-Event Audio into 3D Scenes
}

\author{Shuaihang Yuan, Congcong Wen, Muhammad Shafique, Anthony Tzes and Yi Fang
}

\begin{document}

\maketitle
\thispagestyle{empty}
\pagestyle{empty}

\begin{abstract}

The rapid advances in audio analysis underscore its vast potential for human–computer interaction, environmental monitoring, and public safety; yet, existing audio‐only datasets often lack spatial context. To address this gap, we present two novel audio–spatial scene datasets, Audio-ScanNet and Audio-RoboTHOR, designed to explore audio-conditioned tasks within 3D environments. By integrating audio clips with spatially aligned 3D scenes, our datasets enable research on how audio signals interact with spatial context. To associate audio events with corresponding spatial information, we leverage the common sense reasoning ability of large language models and supplement them with rigorous human verification, This approach offers greater scalability compared to purely manual annotation while maintaining high standards of accuracy, completeness, and diversity, quantified through inter‐annotator agreement and performance on two benchmark tasks: audio‐based 3D visual grounding and audio‐based robotic zero‐shot navigation. The results highlight the limitations of current audio‐centric methods and underscore the practical challenges and significance of our datasets in advancing audio‐guided spatial learning.


\end{abstract}

\section{INTRODUCTION}
\label{sec:intro}

In recent years, the emerging field of audio analysis has garnered widespread attention, highlighting its substantial potential and broad applicability across diverse disciplines. As a fundamental element of human experience, sound offers an abundant source of information, enhancing our comprehension of the world. Research in audio analysis has evolved, extending from augmenting human-computer interaction to advancing environmental monitoring and public safety initiatives. The ability to precisely capture, interpret, and leverage audio signals is paramount for the development of intelligent systems that interact and adapt to their environments in real time. Key tasks in this domain include speech recognition~\cite{weng2023deep, chen2022noise}, acoustic scene classification~\cite{qu2022acoustic, hasan2022genetic}, and sound event detection~\cite{mnasri2022anomalous, nguyen2022salsa}. These tasks are predominantly conducted on audio-event datasets, with prominent examples including ESC~\cite{piczak2015esc}, Audio Set~\cite{gemmeke2017audio}, and FSD~\cite{fonseca2021fsd50k}
. These datasets, primarily characterized by manually annotated audio categorizations, are invaluable for sound recognition and analysis, providing extensive resources for the comprehension and classification of a wide array of sound events. However, they inherently lack visual context, a crucial element in many real-world scenarios.

The integration of visual information with audio substantially enhances event recognition by providing supplementary cues that help disambiguate sound sources. However, most existing audio-visual datasets~\cite{chen2020vggsound,fela2022perceptual} are confined to images or videos and thus fail to capture the full three-dimensional complexity of real-world environments. This lack of spatial context limits the applicability of current datasets and methods, leaving a notable gap in large-scale 3D scene–audio datasets despite the richer spatial information that 3D data offers. 

To bridge this gap and enable advanced multimodal learning in spatially rich settings, we introduce two audio–spatial datasets that integrate immersive 3D environments with corresponding event audios. Creating such datasets, however, presents several challenges. 
Traditional audio datasets lack the necessary spatial context to accurately map sound events to specific objects, and generating plausible audio events for a diverse range of object categories is non-trivial—requiring both contextual inference and the reconciliation of varying labeling conventions across audio-event datasets. Moreover, ensuring high-quality data demands a combination of the strong reasoning capabilities of large language models for event generation and rigorous manual verification, a process that raises scalability concerns. To address these challenges, we propose a novel method that leverages GPT-4 to generate object–event pairs and establish a unified event mapping across multiple audio datasets. This automated process, augmented by targeted manual review, culminates in accurately mapping audio back to the spatial coordinates of objects within 3D scenes. Specifically, we selected the ScanNet dataset~\cite{dai2017scannet} from 3D computer vision and the RoboTHOR dataset~\cite{deitke2020robothor} from robot navigation as our foundational scene data, and we incorporated audio from the ESC~\cite{piczak2015esc}, FSD50K~\cite{fonseca2021fsd50k}, EPIC-SOUNDS~\cite{huh2023epic}, and ReaLISED~\cite{mohino2022introducing} datasets. Experiments on audio-spatial grounding and navigation reveal that current methods relying on audio inputs have limitations, underscoring both the challenges addressed by our approach and its potential to advance multimodal model learning in realistic environments. The main contributions of this paper are summarized as:

\begin{itemize}
    \item We propose two audio-scene datasets, Audio-ScanNet and Audio-RoboTHOR, which combine audio clips with 3D visual scenes, and offer a more comprehensive and realistic platform for multimodal learning. The proposed datasets are pivotal in exploring how audio interacts with and complements 3D visual scenes, providing new insights into spatial audio-visual relationships.

    \item We introduce a novel approach to dataset creation, combining the prowess of advanced large language models like GPT-4 with meticulous manual verification. This approach not only ensures high-quality dataset creation but also paves the way for future innovations in similar domains.

    \item We designe two benchmark tasks focused on audio-based 3D visual grounding and audio-based robotic navigation. These benchmarks serve as critical tools for evaluating the effectiveness of our proposed datasets. The results of these tasks do not merely attest to the utility of Audio-Scannet and Audio-RoboTHOR but also shed light on potential avenues for further advancement in the realm of multimodal learning.

\end{itemize}

\section{RELATED WORK}
\label{sec:rw}

\subsection{Audio Dataset}
In addressing the challenges associated with sound event recognition, the academic community has seen the emergence of several audio datasets~\cite{piczak2015esc, gemmeke2017audio, chen2020vggsound, chen2020soundspaces, chen2022soundspaces, huh2023epic, mei2023wavcaps} in recent years. A significant early contribution was made by Piczak et al.~\cite{piczak2015esc}, who compiled a new annotated collection for environmental sound classification from the Freesound project. This collection comprises 2,000 short audio clips, categorized into 50 classes, each representing a different common sound event. Besides, Gemmeke et al.~\cite{gemmeke2017audio} introduced the Audio Set, a comprehensive dataset that encompasses a manually-annotated catalog of 632 audio event categories, systematically arranged in a hierarchical structure with a maximum depth of six levels. Building upon Audio Set, Fonseca et al.~\cite{fonseca2021fsd50k} developed the FSD50K dataset. This dataset includes over 51,000 audio clips, collectively amounting to more than 100 hours of audio. Each clip in FSD50K is manually labeled across 200 classes, derived from the comprehensive AudioSet Ontology. More recently, EPIC-SOUNDS~\cite{huh2023epic} was proposed by Huh et al. by capturing temporal extents and class labels within the audio stream of the egocentric videos. EPIC-SOUNDS includes 78.4k categorized segments that encompass a diverse range of audible events and actions. These segments are systematically classified across 44 distinct classes. In addition to these categorized segments, the dataset also includes 39.2k segments that are not categorized.

\subsection{Audio-Visual Learning}

Recent advancements in audio-visual learning have substantially progressed the field, bringing forth a range of influential and innovative developments. Key tasks in this domain include Audio-Visual Sound Separation \cite{Ephrat_2018,
gao2019coseparating,
xu2019recursive,
zhao2019sound,
rouditchenko2019selfsupervised,
gan2020music}, Audio-Visual Sound Synthesis\cite{oord2016wavenet,
richard2021neural,
donahue2019adversarial,
engel2019gansynth,
chen2022visual,
gao20192,
zhou2020sepstereo}, and Audio-Visual Video Understanding \cite{gao2019listen,
tian2020unified,
zhou2020sep,
zhou2021pose,
xu2021visually}  
, each contributing uniquely to the advancement of multimedia technologies. In this paper, we focus our investigation on two tasks that are particularly relevant to 3D scenarios: Audio-Visual Grounding \cite{
hu2020cross,
hu2020curriculum,
afouras2020self,
qian2020multiple,
hu2020discriminative} and Audio-Visual Navigation\cite{chen2020soundspaces}. Audio-Visual Grounding involves identifying visual objects that correspond to specific sounds within everyday audiovisual environments, providing a more integrated and context-aware understanding of multimedia content. Audio-visual navigation, on the other hand, is concerned with the development of autonomous systems or robots capable of navigating environments by utilizing both auditory and visual inputs, thereby enhancing their spatial awareness and interaction with their surroundings. This dual focus on grounding and navigation underscores the evolving nature of audio-visual learning and its increasing applicability to practical, real-world applications.




\section{AUDIO-SCENE DATASET}
\label{sec:approach}

\begin{figure*}[t]
    \centering
    \includegraphics[width=1.0\textwidth]{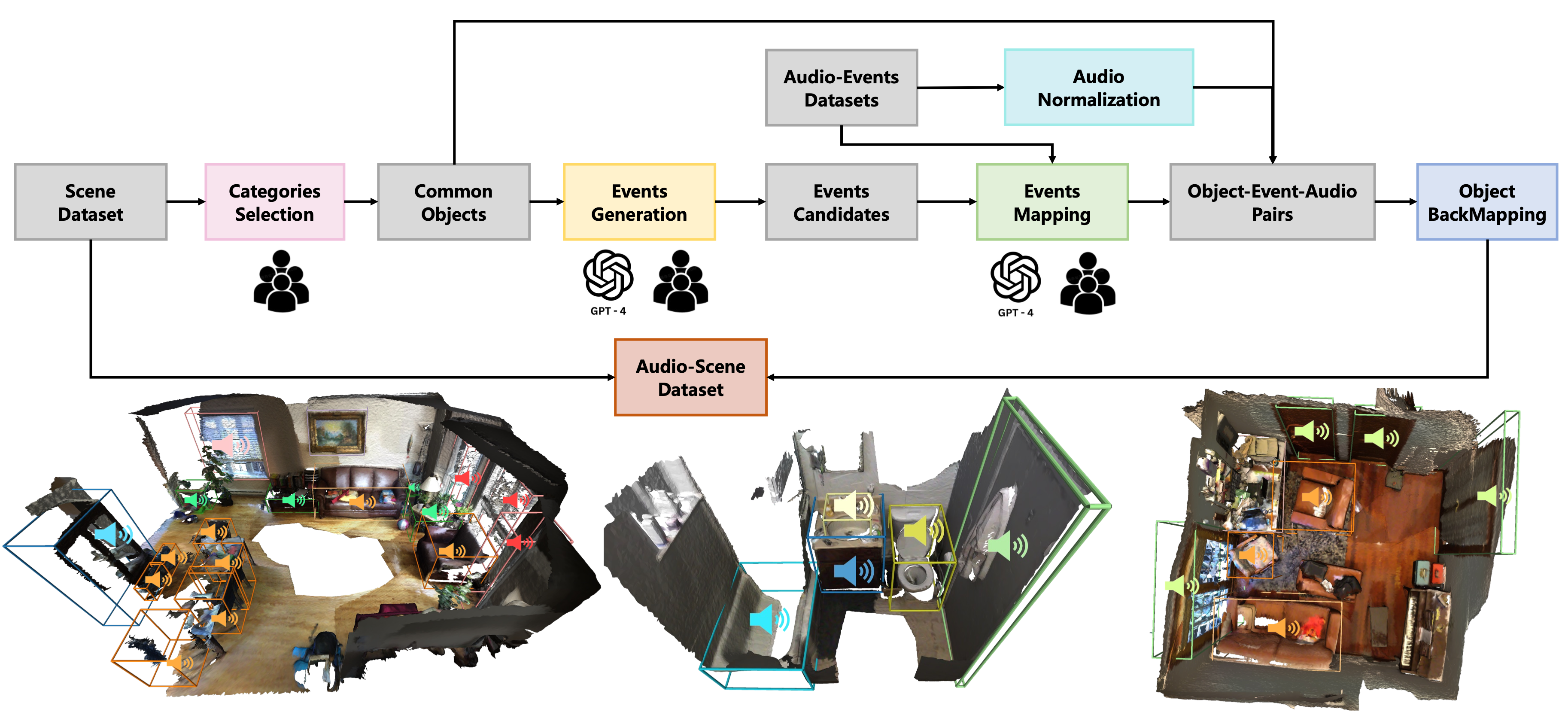} 
    \vspace{-8mm}
    \caption{Illustration of the process of Audio-Scene dataset creation. It involves selecting categories from a scene dataset to identify common objects, followed by a GPT-4 facilitated event generation phase producing corresponding audio events. Concurrently, audio events datasets are normalized. These events are then mapped to scene objects using GPT-4, forming object-event-audio pairs. The pairs are backmapped into the original 3D scene, finalizing the dataset. } 
    \label{fig:dataset_procedure}
    \vspace{-5mm}
\end{figure*}

\subsection{Overall Procedure}
The Overall Procedure of Audio-Scene Dataset is illustrated in Figure~\ref{fig:dataset_procedure}. We first identified common categories in the ScanNet and RoboTHOR datasets, ultimately selecting 36 categories including alarmclock, apple and others for our study. All detailed object categories can be found in Supplementary. Subsequently, we designed a object-event generation(OEG) module based on large language models (LLMs) to infer potential events associated with these categories, and we meticulously verified the plausibility of these events through manual review. Following this, we matched these events with those in four prevalent audio-event datasets, namely ESC, FSD, EPIC, and REAL. Using our specially designed matching module, we associated the events from the existing datasets with the events we had previously generated, thereby establishing a connection between objects and their corresponding audio. Finally, we mapped the object-related audio back to the respective scenes, resulting in the creation of two audio-scene datasets: Audio-ScanNet and Audio-RoboTHOR. 


\subsection{Data Sources}

\subsubsection{Scene Data Sources}

\noindent\textbf{ScanNet} \cite{dai2017scannet} dataset emerges as a comprehensive solution to the scarcity of large, labeled datasets in RGB-D scene understanding. Encompassing over 2.5 million views from 1513 indoor scenes, it provides an unprecedented depth of data, including 3D camera poses, surface reconstructions, and semantic segmentations. The innovation behind ScanNet lies in its user-friendly RGB-D capture system, which streamlines both automated surface reconstruction and the crowdsourcing of semantic annotations. This rich dataset facilitates cutting-edge performance in several 3D scene understanding tasks, such as 3D object classification, semantic voxel labeling, and CAD model retrieval. 

\noindent\textbf{RoboTHOR}\cite{deitke2020robothor}
is developed by the Allen Institute for AI, represents a groundbreaking platform in artificial intelligence and robotics research, merging simulated and physical environments to address key challenges in AI model transferability. The dataset collected from physical space with 8.8m by 3.9m comprises a set of 89 meticulously designed apartments, segmented into 75 for training and validation, 4 in test-dev for real-world validation, and 10 in test-standard for blind physical tests.  And the RoboTHOR environments include 11 types of furniture, like TV stands and dining tables, and 32 types of smaller objects, such as mugs and laptops. These scenes are meticulously designed to provide a rich tapestry of domestic environments, each with unique configurations and object placements, offering a comprehensive testing ground for AI models.

\subsubsection{Audio-Event Data Sources}
\noindent\textbf{Environmental Sound Classification (ESC) \cite{piczak2015esc}} is divided into three distinct segments: ESC-50, ESC-10, and ESC-US. The core component, ESC-50, comprises 2,000 meticulously labeled environmental recordings. These recordings are uniformly distributed across 50 categories, resulting in 40 distinct audio clips per category. The categories themselves are organized into five broadly defined major groups, each encompassing ten classes. 

\noindent\textbf{Freesound Dataset 50k (FSD50K)~\cite{fonseca2021fsd50k}} encompasses a total of 51,197 audio clips, collectively amounting to over 100 hours of audio. Each audio clip within FSD50K has been meticulously labeled, utilizing a set of 200 classes derived from the AudioSet Ontology. The organizational structure of these classes in FSD50K is hierarchical, comprising 144 leaf nodes and 56 intermediate nodes. Sourced from Freesound, the contents of FSD50K are licensed under various Creative Commons (CC) licenses.  


\noindent\textbf{EPIC-SOUNDS~\cite{huh2023epic}} comprises a total of 78,366 categorized segments. These categorized segments span a diverse range of audible events and actions, systematically distributed across 44 distinct classes. Each segment is a temporal annotation with an average duration of 4.9 seconds, offering a detailed insight into various soundscapes.

\noindent\textbf{Real-Life Indoor Sound Event Dataset (ReaLISED) ~\cite{mohino2022introducing}} comprises 2,479 clips of isolated sounds collected from total 3,624.51 seconds of audio. This dataset encompasses a diverse range of eighteen (18) distinct sound event classes, each carefully chosen to represent the dynamic acoustic environment of a typical household.


\subsection{Data acquisition}


\subsubsection{Categories Selection}
Given two scene datasets, ScanNet and RoboTHOR, each scene $S_i$ of dataset contains many objects $ \{o_1, o_2, \ldots, o_n\} $, where each $o_i$ signifies an independent object within the scene. We identify a set of common categories from two datasets, denoted as \( C = \{c_1, c_2, \ldots, c_{m}\} \), which includes various objects like beds, alarm clocks, etc. 

\subsubsection{Events Generation from Object Categories}
For each category \( c_i \in C \), we design a LLM-based Object-Event Generation (LLM-OEG) module to infer a set of potential events \( e_i \). Our approach is primarily inspired by the remarkable success of large language models (LLMs) in a variety of fields, particularly in natural language processing and understanding. Specifically, we utilize the GPT-4 model, a state-of-the-art LLM known for its exceptional ability to generate human-like responses to natural language prompts. GPT-4's capabilities in understanding, reasoning, and dialogue have been widely recognized and acclaimed in recent research, making it an ideal choice for our module. Utilizing GPT-4, we aim to effectively generate a set of plausible events associated with each object category. This is achieved by crafting appropriate prompts that guide the model to produce relevant and contextually accurate event descriptions. The strength of GPT-4 in generating coherent and context-aware textual content significantly enhances our ability to obtain a diverse range of events that are intricately linked to the specified object categories. Moreover, this methodology has the potential to markedly reduce the discard rate of raw descriptions often encountered in traditional approaches, thereby enhancing the overall quality and usability of the generated captions. 

\begin{figure*}[ht]
    \centering
    \includegraphics[width=0.95\textwidth]{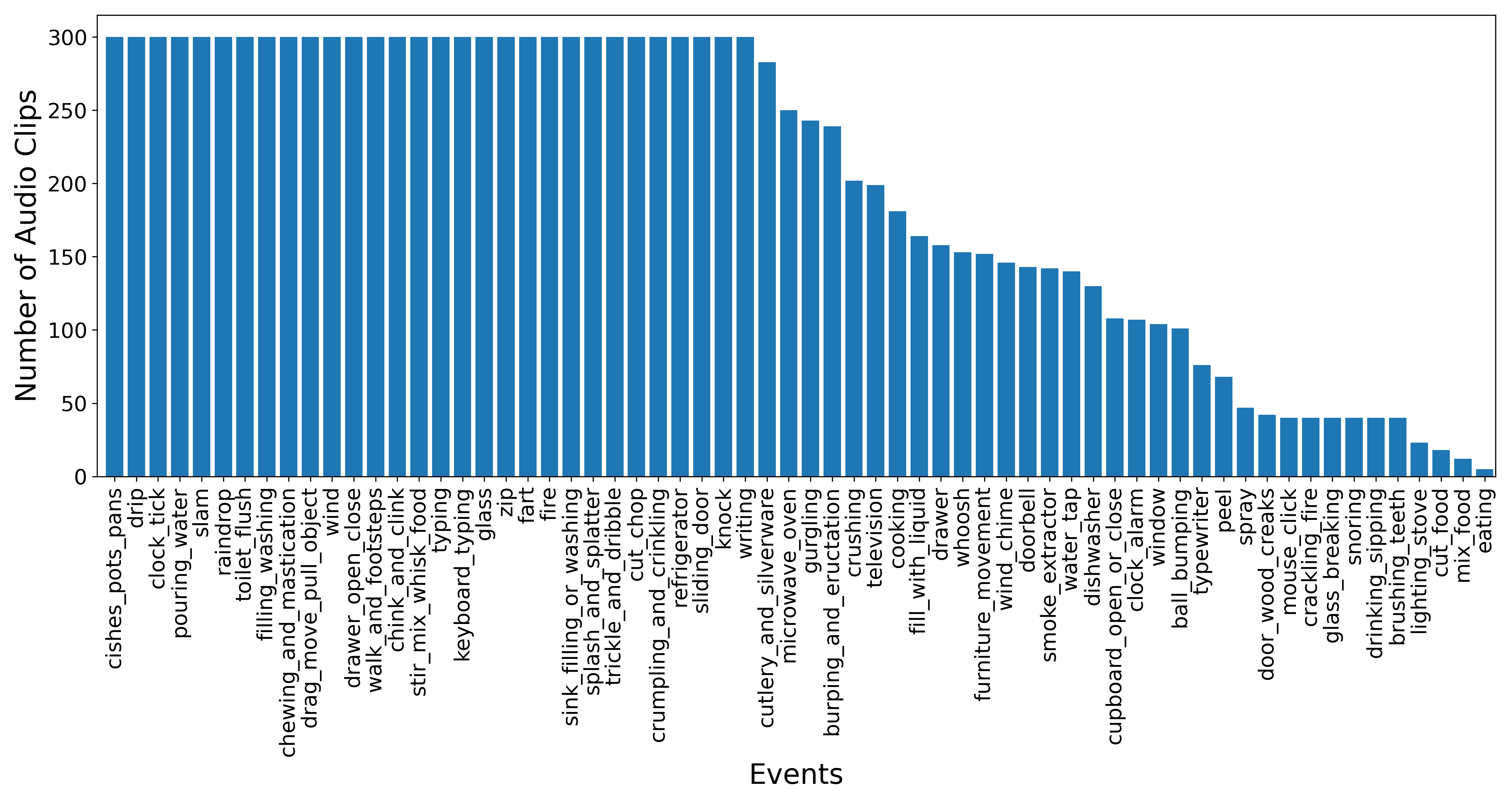} 
    \vspace{-7mm}
    \caption{Distribution of audio clips across different events in our dataset.}
    \label{fig:dist_event}
    \vspace{-5mm}
\end{figure*}

To ensure the reliability and applicability of these events, we implement a rigorous post-generation review process. Each event generated by GPT-4 is meticulously evaluated manually to assess its reasonableness and relevance to the corresponding object category. This manual verification serves as a crucial step in ensuring the quality and accuracy of our LLM-OEG module's outputs, thereby contributing to the robustness of our research methodology. The number of audio clips for different events can be found in Figure \ref{fig:dist_event}

\subsubsection{Events Mapping}

In our study, we define \( D_{ESC}, D_{FSD}, D_{EPIC}, \) and \( D_{REAL} \) as four prominent audio-event datasets. The union of events from these datasets is represented as \( E = E_{ESC} \cup E_{FSD} \cup E_{EPIC} \cup E_{REAL} \), encompassing the entire spectrum of audio events available in our study. The primary objective is to develop a matching function, denoted as \( f: e \rightarrow E \), which systematically links each inferred event \( e_i \) from our dataset to the corresponding events in these established audio-event datasets. This function is crucial as it forms a foundational basis for correlating the objects in set \( C \) with their respective audio events in these datasets, thereby facilitating a multi-modal analysis.

It should be noted that due to varying standards in dataset creation, there inevitably exist discrepancies in event representations across different datasets. For instance, the same event might be labeled as “clock\_alarm” in one dataset and simply as “alarm” in another. To address this challenge, we initiate our process by consolidating the events from the original four data sources using the GPT-4 model. This consolidation step involves merging semantically similar events under a unified event name, effectively standardizing the event nomenclature across datasets.


To enhance the accuracy and relevance of our matching process, we employ the GPT-4 model to generate a score matrix \( M_{GPT4} \). This matrix, with scores ranging from 1 to 5, quantifies the relevance between each event \( e_i \) in our dataset and the events \( E_i \) in the standard datasets. A higher score indicates greater relevance, thus providing a measurable metric for event matching. Concurrently, to ensure the reliability and validity of our automated scoring, we also create a manually evaluated score matrix \( M_{Human} \). This manual scoring serves as a critical benchmark to validate the automated scores generated by GPT-4.

In the final phase of our methodology, we focus on selecting the most relevant events for mapping. Events with scores exceeding a threshold of 3 in both \( M_{GPT4} \) and \( M_{Human} \) are shortlisted, ensuring that our selection process is robust and takes into account both automated and human assessments. This dual-matrix approach enables us to identify and select the most pertinent events for establishing audio-object mappings, which are integral to our research in audio-scene synthesis and analysis.

\subsubsection{Mapping Audio to Scene}

Upon establishing the mapping relationship between the events \( e \), generated based on object categories, and the events \( E \) in the existing audio-event datasets, we proceed to construct a mapping between the object categories \( C \) and the audio \( A \) from these source datasets. This is achieved through indexing, which allows us to associate each object category with its corresponding audio. Furthermore, we refine this process by mapping the audio \( A \) back to the spatial coordinates of each object within the scene. This spatial mapping is a critical step, as it ensures that the audio is accurately aligned with the physical location of the objects in the scene. By doing so, we enhance the realism and contextual relevance of our synthesized audio-scene dataset, making it a more effective tool for applications that require a detailed and accurate representation of audio-visual environments.

\subsection{Dataset Analysis}

\begin{figure}[h]
    \centering
    \includegraphics[width=1\linewidth]{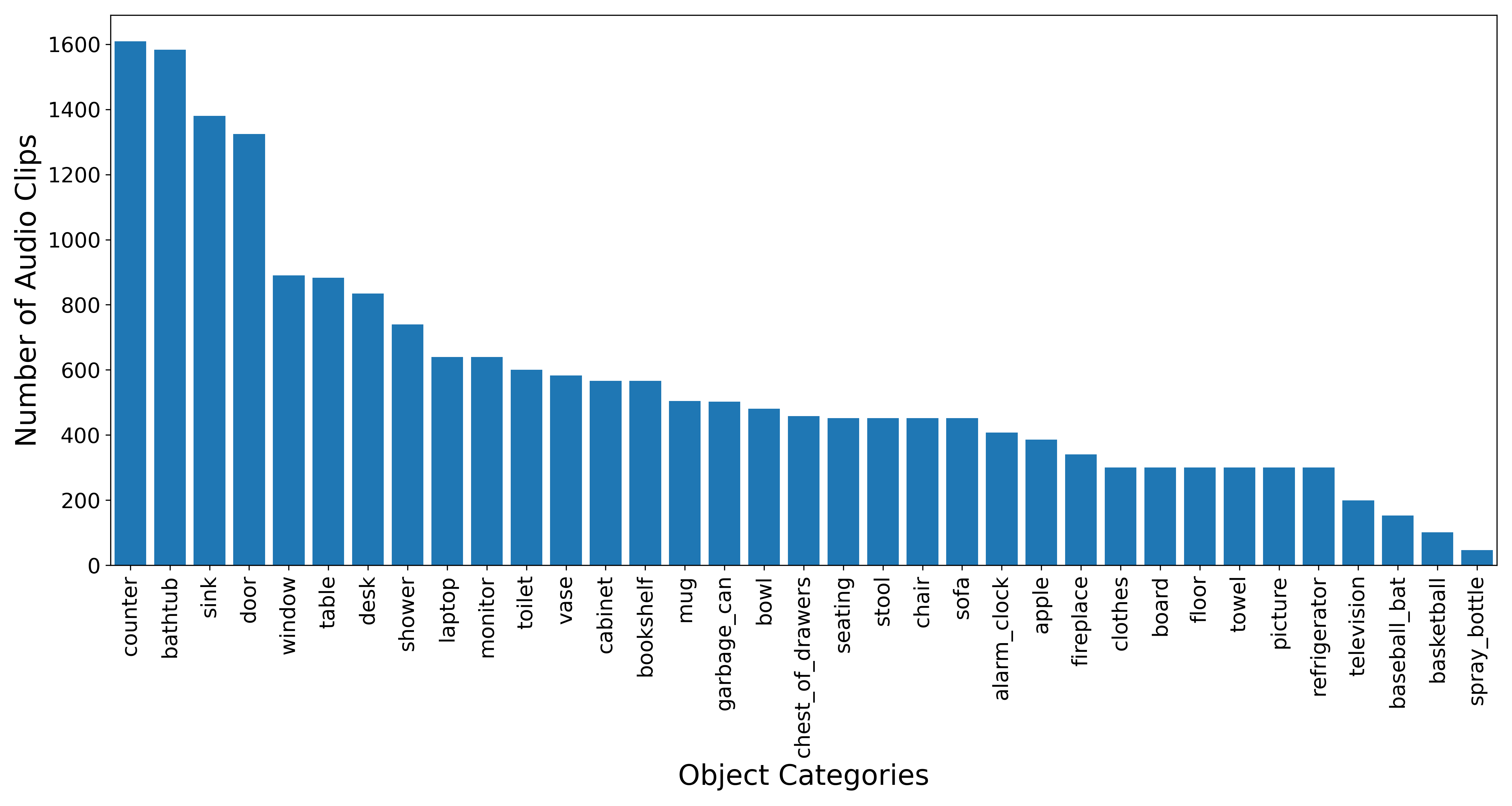}
    \vspace{-10mm}
    \caption{Data distribution of audio clips by object categories.}
    \vspace{-5mm}
    \label{fig:dis_cat}
\end{figure}

Our dataset encompasses an extensive collection of 12,876 audio clips, which are derived from 65 distinct events and correspond to 36 common object categories. We have conducted a thorough statistical analysis of the audio clip distribution, categorized by event type and object category. The results of this analysis are presented in Figures 1 and 2. As illustrated in Figure 1, each event category encompasses up to 300 audio clips as shown in Figure \ref{fig:dist_event}. Notably, thirty events are represented by 300 clips each, and fifty events have more than 100 clips, as shown in Figure \ref{fig:dis_cat}. On average, each event category comprises approximately 201 audio clips. Similarly, Figure 2 provides insights into the distribution of audio clips across different object categories. The ``counter'' category exhibits the highest number of clips, amounting to 1,600. Furthermore, we observed that 89\% of the object categories are associated with more than 200 audio clips. On average, each object category is represented by 358 clips. This detailed enumeration and analysis of the audio clips not only demonstrate the comprehensive nature of our dataset but also provide valuable insights into the diversity and richness of the soundscapes associated with each event and object category. 


\section{BENCHMARK}
\label{sec:exp}

\subsection{3D Visual Grounding}

\subsubsection{Baselines}

\noindent\textbf{3DVG-Transformer}\cite{zhao2021_3DVG_Transformer} is designed for the task of text-based 3D object localization. This method leverages a coordinate-guided contextual aggregation module for enhanced proposal generation. It also incorporates a multiplex attention module during the cross-modal feature fusion stage to refine the selection of the target objects.  

\noindent\textbf{3D-SPS}\cite{Luo_2022} addresses the challenge of localizing a target object in a 3D space using only a textual description. The core of 3D-SPS is its Referred Point Progressive Selection (RPPS) mechanism, which efficiently narrows down the search space in 3D point clouds by progressively selecting points that are more likely to correspond to the described object. Unlike traditional methods that often require multi-stage processing and are computationally intensive, 3D-SPS offers a more streamlined and efficient approach. 

\noindent\textbf{Multi-View Transformer}\cite{huang2022multi} addresses projecting the 3D scene into a multi-view space, capturing position information from various angles and integrating these perspectives to form a view-agnostic representation. 

\subsubsection{Baseline Adaptation}
In adapting the three baseline methodologies, originally configured for text-based 3D object localization, we have implemented a substitution of the text encoder component with an audio encoder. We have selected the wav2clip \cite{wu2022wav2clip} model to function as our new audio encoder while maintaining the integrity of the remaining elements of the baseline architectures unchanged. To formally describe this modification, let \( E_{text} \) represent the original text encoder and \( E_{audio} \) denote the wav2clip audio encoder that we have introduced. The encoding process of audio is defined as:
 $V_{audio} = E_{audio}(A)$
where \( A \) signifies the input audio data and \( V_{audio} \) is the resultant audio feature vector. By adopting this encoder replacement, the baseline methods are transformed to utilize audio cues for 3D object localization.

\subsubsection{The Proposed Method}
We propose a method to improve audio-based 3D object localization. Initially, our method utilizes an audio-to-event recognition process, converting audio data into events using the wav2clip \cite{wu2022wav2clip} model as the audio encoder. Subsequent to event recognition, the VoteNet \cite{qi2019deep} object detection algorithm is employed to analyze and produce the object detection of the 3D scene. Building on this, we integrate the LLaMA \cite{touvron2023llama} language model to establish a probabilistic link between the recognized events and the spatially localized objects. The LLaMA model processes the event text and the object detection results to infer which object is most likely responsible for the audio event. It outputs a classification label indicating the object within the scene with the highest likelihood of being the event's source. This method is carefully designed to refine the accuracy of object localization in 3D environments by leveraging the distinctive signatures of audio events. We use mAP@mIoU  as the evaluation metric, where mIoU $\in$ \{0.25, 0.5\}. mAP@mIoU measures the fraction of language queries whose predicted box overlaps the ground truth box with 3D intersection over the union (IoU) higher than mIoU.

\begin{figure*}[h]
    \centering
    \includegraphics[width=.9\textwidth]{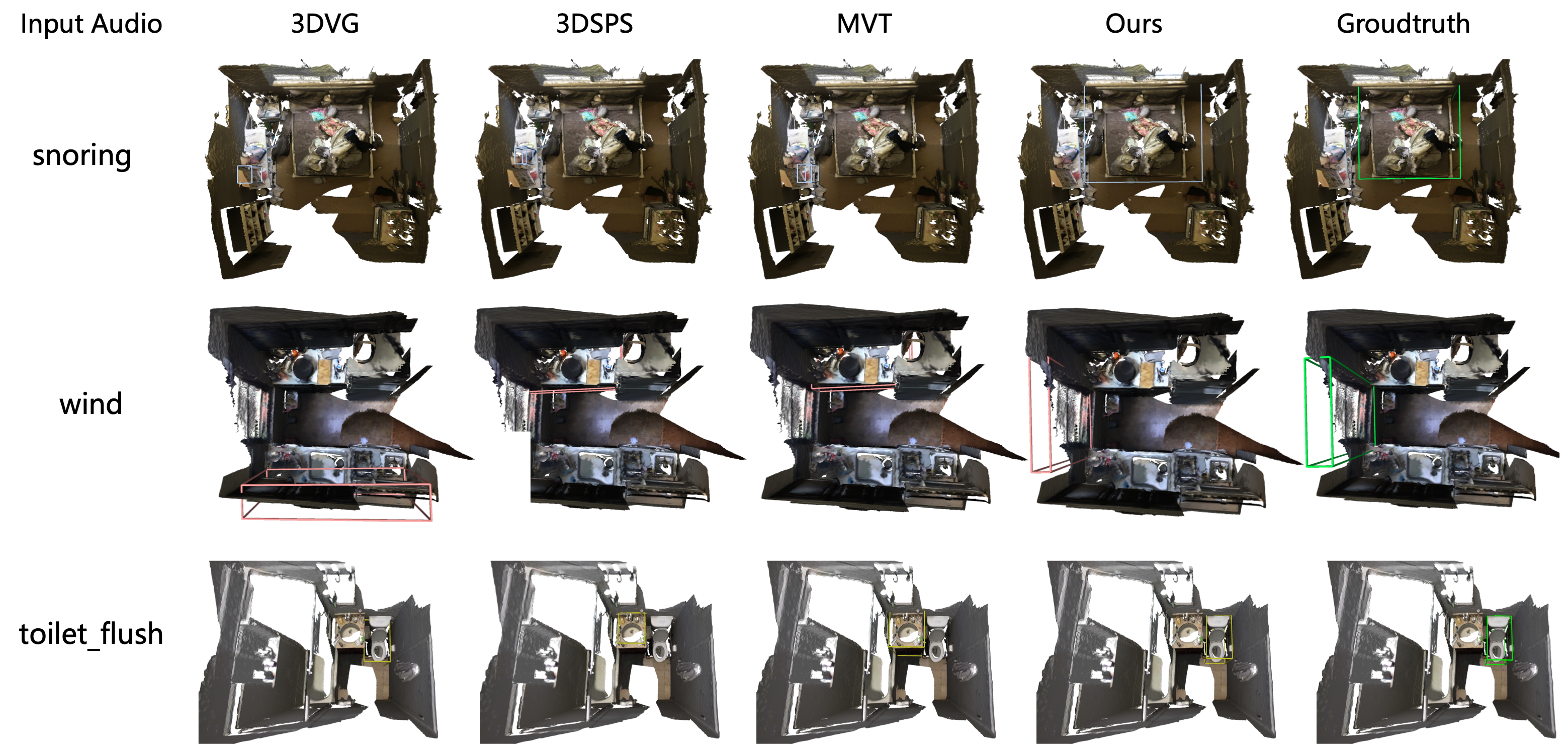}

    \caption{Visualization of Audio-based 3D Object Grounding Methods on the ScanNet Dataset. The figure illustrates the localization results for different audio inputs: 'snoring,' 'wind,' and 'toilet flush,' using various methods against the ground truth.}
    \label{fig:scannet}

\end{figure*}

\begin{table*}[h]
\centering
\caption{Performance Comparison of Audio-Based Object Grounding Methods on the ScanNet Dataset.}
\resizebox{0.8\textwidth}{!}{
\begin{tabular}{lcccccccc}
\toprule
Object Category & \multicolumn{2}{c}{3DV-TRANS\cite{zhao2021_3DVG_Transformer}} & \multicolumn{2}{c}{3D-SPS\cite{Luo_2022}} & \multicolumn{2}{c}{MVT\cite{huang2022multi}} & \multicolumn{2}{c}{Ours} \\
\cmidrule(lr){2-3} \cmidrule(lr){4-5} \cmidrule(lr){6-7} \cmidrule(lr){8-9}
                & AP @ 0.25       & AP @ 0.5      & AP @ 0.25     & AP @ 0.5    & AP @ 0.25    & AP @ 0.5  & AP @ 0.25    & AP @ 0.5 \\
\midrule
Bathtub        & 0.373 & 0.095 & 0.001 & 0.001 & 0.471 & 0.449 & \textbf{0.636} &  \textbf{0.629} \\
Bed            & 0.011 & 0.010 & 0.000 & 0.000 & \textbf{0.972} & \textbf{0.944} & 0.136 &  0.136 \\
Bookshelf      & 0.198 & 0.038 & 0.190 & 0.134 & 0.215 & 0.178 & \textbf{0.634} &  \textbf{0.624} \\
Cabinet        & 0.005 & 0.001 & 0.015 & 0.010 & 0.159 & 0.102 & \textbf{0.422} &  \textbf{0.384} \\
Chair          & 0.081 & 0.035 & 0.600 & 0.508 & 0.156 & 0.113 & \textbf{0.901} &  \textbf{0.800} \\
Counter        & 0.000 & 0.000 & 0.371 & 0.200 & 0.328 & 0.172 & \textbf{0.434} &  \textbf{0.344} \\
Desk           & 0.205 & 0.047 & 0.349 & 0.259 & \textbf{0.256} & 0.202 & 0.215 &  \textbf{0.204} \\
Door           & 0.000 & 0.000 & 0.000 & 0.000 & \textbf{0.355} & 0.154 & 0.327 & \textbf{0.282} \\
Picture        & 0.000 & 0.000 & 0.000 & 0.000 & \textbf{0.188} & \textbf{0.047} & 0.014 &  0.014 \\
Refrigerator   & 0.000 & 0.000 & 0.000 & 0.000 & \textbf{0.643} & \textbf{0.486} & 0.242 &  0.242 \\
Shower Curtain & 0.127 & 0.027 & 0.247 & 0.181 & 0.312 & 0.263 & \textbf{0.860} &  \textbf{0.793} \\
Sink           & 0.137 & 0.025 & 0.818 & 0.352 & 0.306 & 0.152 & \textbf{0.619} &  \textbf{0.472} \\
Sofa           & 0.096 & 0.029 & 0.298 & 0.298 & 0.146 & 0.127 & \textbf{0.298} &  \textbf{0.298} \\
Table          & 0.077 & 0.018 & 0.037 & 0.030 & 0.164 & 0.108 & \textbf{0.498} &  \textbf{0.465} \\
Toilet         & 0.189 & 0.089 & 0.169 & 0.169 & 0.760 & 0.715 & \textbf{0.795} &  \textbf{0.795} \\
Window         & 0.018 & 0.001 & 0.039 & 0.018 & 0.334 & 0.173 & \textbf{0.698} &  \textbf{0.551} \\
\hline
mAP           & 0.084 & 0.026 & 0.196 & 0.180 & 0.360 & 0.274 & \textbf{0.429} & \textbf{0.391} \\
\bottomrule
\end{tabular}}
\label{tab:scannet}

\end{table*}

\subsubsection{Results}
Table \ref{tab:scannet} presents the performance of our audio-based localization method alongside various baseline approaches. The results indicate that our method surpasses the baseline in terms of accuracy. Specifically, there is an improvement of approximately 19\% in mAP@0.25 and 42\% in mAP@0.5. Additionally, Figure \ref{fig:scannet} provides a visual representation of the results, demonstrating that our approach accurately identifies the audio event and correctly infers the object most likely to produce the sound.

\subsection{Zero-Shot Object Navigation}

\begin{figure}[h]
    \centering
    \includegraphics[width=.9\linewidth]{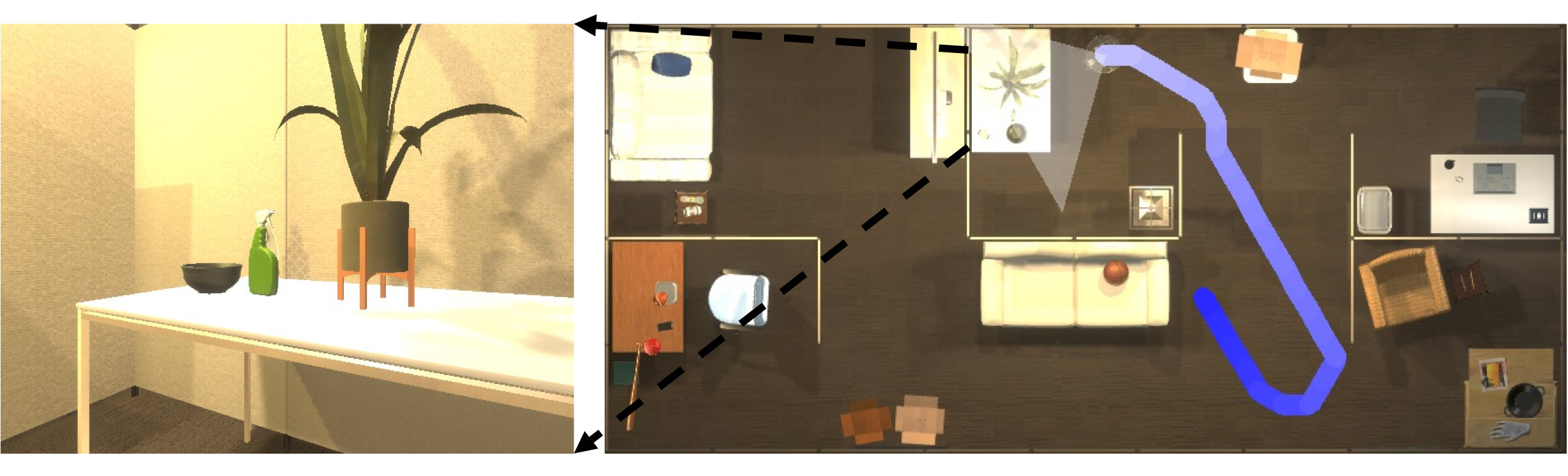}

    \caption{Robot Navigation Path Triggered by Audio Cue. The right side shows trajectory in response to the sound of a plant, and the left side displays robot's view of the target object upon arrival.}
    \label{fig:nav}
    \vspace{-3mm}
\end{figure}

\subsubsection{Baselines}
\noindent \textbf{CLIP-Ref~\cite{radford2021learning}:}
CLIP \cite{radford2021learning} with k referring expressions. The process involves maching a CLIP-derived visual representation of the current observation with k different CLIP textual embeddings, each delineating possible locations of the target object. 


\noindent \textbf{CLIP-Patch~\cite{radford2021learning}:}
CLIP \cite{radford2021learning} with k image patches \cite{gadre2022cow}. It discretizes the image into k smaller patches. Each patch undergoes analysis via the CLIP vision backbone to extract patch features. 

\noindent \textbf{CLIP-Grad~\cite{radford2021learning}:}
CLIP with gradient relevance \cite{Chefer_2021_CVPR}. This method involves leveraging a target CLIP text embedding alongside gradient information derived from the CLIP vision backbone to generate a relevance map across image pixels that qualitatively segments the target object. 


\noindent \textbf{MDETR~\cite{kamath2021mdetr}:}
The model enhances the capabilities of the DETR detector \cite{carion2020endtoend} by augmenting its functionality to simultaneously process text and images, thereby facilitating the generation of bounding box detections. In our work, we utilize the MDETR model specifically for extracting object relevance in relation to designated targets.


\noindent \textbf{CLIP-OWL~\cite{gadre2022cow}:}
OWL-ViT detection \cite{minderer2022simple} explore a method for transforming CLIP-like models into object detectors through fine-tuning on a set prediction task. We employ OWL-ViT for directly querying images in search of specified targets.

\noindent \textbf{ESC~\cite{zhou2023esc}:}
It uses a grounded language-image model (GLIP) to detect objects and rooms in the environment, generating a semantic map for navigation. With the aid of the semantic map, ESC performs reasoning based on large language models to rate the likelihoold of each frontier location based on the goal object's relation to other objects and rooms. 


\subsubsection{Baseline Adaptation}
Above baseline methodologies employed in Audio Navigation are derived from pre-existing models designed for Zero-Shot Object Navigation (ZSON) tasks, specifically tailored for the RoboTHOR dataset \cite{deitke2020robothor}. In the standard ZSON paradigm, autonomous agents are provided with a textual prompt specifying the name of the target object. This research extends the scope to incorporate audio navigation, wherein the conventional target input, typically a text prompt, is substituted with an audio clip representing the target object. To ensure minimal alterations to the original models, audio classification network based on wav2clip\cite{wu2022wav2clip} has been integrated into each baseline model. 
For the classification of input audio into distinct object categories, we firstly adapt the audio embedding mechanism from wav2clip \cite{wu2022wav2clip} to process raw audio files. This audio embedder initially adjusts the length of the audio files to a standardized duration, either by padding or trimming, and normalizes the audio values. Subsequently, the processed audio is input into an encoder, which generates a latent tensor representation. The shape of this tensor is contingent upon the channel configuration of the original audio file: it yields a tensor of dimensions $[1, 512, 512]$ for solo-channeled audio and $[2, 512]$ for duo-channeled audio. In the latter case, the duo-channeled latent matrix is averaged and transformed to a $[1, 512]$ single-channeled matrix. Then the latent features are streamlined through a flattening layer and subsequently through a MLP (Multi-Layered Perception) network with demensionality reduction of $[512, 384]$ and $[384, k]$, where k represents the total number of object categories to be classified. This network processes audio clips, categorizing them into distinct object classes. Subsequently, the identified object categories from the classification network are fed into the ZSON models to perform the navigation task.

\begin{table}[h]
\centering
\vspace{-3mm}
\caption{The quantitative results for the audio-based object navigation on the RoboTHOR dataset.}
\begin{tabular}{ l  l  l }
\toprule
\multirow{2}{*}{Model}  & \multicolumn{2}{l}{Performance} \\

 & SPL & SR \\
\midrule
CLIP-Ref.\cite{gadre2022cow} & 1.0 & 1.8 \\

CLIP-Patch.\cite{gadre2022cow} & 7.7 & 15.3 \\

CLIP-Grad.\cite{gadre2022cow} & 7.4 & 12.1 \\

MDETR.\cite{gadre2022cow}& 0.0 & 0.0 \\

CLIP-OWL. \cite{gadre2022cow}& 13.4 & 21.9 \\

ESC\cite{zhou2023esc} & \textbf{14.1 }&\textbf{ 27.1} \\

\bottomrule

\end{tabular}
\label{tab:robothor}
\vspace{-3mm}
\end{table}

\subsubsection{Object Navigation Metrics}
To measure performance, we employ two widely recognized metrics in the field of object navigation. \textbf{SR}: represents the proportion of episodes in which the agent successfully executes the STOP command within a 1.0 meter radius of the target object. \textbf{SPL}: is calculated by weighting the success rate by the ratio of the oracle shortest path length to the actual path traversed by the agent.

\subsubsection{Results}
Table \ref{tab:robothor} presents the performance of different object navigation baseline baseline approaches. In summary, the \textbf{ESC} model demonstrates the best performance in terms of both navigating efficiently (as indicated by an SPL of 14.1) and successfully reaching the target (as reflected in an SR of 27.1). This suggests the effectiveness of its underlying methodology of commonsense reasoning and scene interpretation capabilities, in comparison to the other models. Following up, the \textbf{CLIP-OWL} model also shows robust performance in navigating to the target object efficiently and accurately with an SPL of 13.4 and an SR of 21.9. \textbf{CLIP-Patch}: and \textbf{CLIP-Grad} shows a lower performance in both SR and SPL, suggesting less accurate scene understanding in in object navigation. On the other hand, \textbf{CLip-Ref} and \textbf{MDETR} models has the lowest performance among all baseline models, suggesting limited effectiveness in the audio navigation task. The visualized navigation sample is shown in Figure \ref{fig:nav}

\section{CONCLUSIONS}
\label{sec:con}

This paper research presents two audio-scene datasets, Audio-ScanNet and Audio-RoboTHOR. These datasets, by encompassing both audio and 3D spatial elements, provide a richer, more comprehensive platform for multimodal learning. In addition, we introduce a novel dataset creation approach of combining large language models like GPT-4 with manual verification, which has not only ensured the quality and realism of our datasets but also opened a new slight for future dataset creation. Moreover, we design two benchmark tasks in audio-based 3D visual grounding and robotic navigation, and validated the effectiveness of the proposed datasets. Our work opens new avenues in audio-visual research, particularly in enhancing real-world applications like augmented reality and smart home systems, and sets the stage for future exploration in multimodal model learning.

{
    \small
    \bibliographystyle{IEEEtran}
    \bibliography{ref}

\begin{thebibliography}{10}
\providecommand{\url}[1]{#1}
\csname url@rmstyle\endcsname
\providecommand{\newblock}{\relax}
\providecommand{\bibinfo}[2]{#2}
\providecommand\BIBentrySTDinterwordspacing{\spaceskip=0pt\relax}
\providecommand\BIBentryALTinterwordstretchfactor{4}
\providecommand\BIBentryALTinterwordspacing{\spaceskip=\fontdimen2\font plus
\BIBentryALTinterwordstretchfactor\fontdimen3\font minus \fontdimen4\font\relax}
\providecommand\BIBforeignlanguage[2]{{%
\expandafter\ifx\csname l@#1\endcsname\relax
\typeout{** WARNING: IEEEtran.bst: No hyphenation pattern has been}%
\typeout{** loaded for the language `#1'. Using the pattern for}%
\typeout{** the default language instead.}%
\else
\language=\csname l@#1\endcsname
\fi
#2}}

\bibitem{weng2023deep}
Z.~Weng, Z.~Qin, X.~Tao, C.~Pan, G.~Liu, and G.~Y. Li, ``Deep learning enabled semantic communications with speech recognition and synthesis,'' \emph{IEEE Transactions on Wireless Communications}, 2023.

\bibitem{chen2022noise}
C.~Chen, N.~Hou, Y.~Hu, S.~Shirol, and E.~S. Chng, ``Noise-robust speech recognition with 10 minutes unparalleled in-domain data,'' in \emph{ICASSP 2022-2022 IEEE International Conference on Acoustics, Speech and Signal Processing (ICASSP)}.\hskip 1em plus 0.5em minus 0.4em\relax IEEE, 2022, pp. 4298--4302.

\bibitem{qu2022acoustic}
Y.~Qu, X.~Li, Z.~Qin, and Q.~Lu, ``Acoustic scene classification based on three-dimensional multi-channel feature-correlated deep learning networks,'' \emph{Scientific Reports}, vol.~12, no.~1, p. 13730, 2022.

\bibitem{hasan2022genetic}
N.~W. Hasan, A.~S. Saudi, M.~I. Khalil, and H.~M. Abbas, ``A genetic algorithm approach to automate architecture design for acoustic scene classification,'' \emph{IEEE Transactions on Evolutionary Computation}, vol.~27, no.~2, pp. 222--236, 2022.

\bibitem{mnasri2022anomalous}
Z.~Mnasri, S.~Rovetta, and F.~Masulli, ``Anomalous sound event detection: A survey of machine learning based methods and applications,'' \emph{Multimedia Tools and Applications}, pp. 1--50, 2022.

\bibitem{nguyen2022salsa}
T.~N.~T. Nguyen, K.~N. Watcharasupat, N.~K. Nguyen, D.~L. Jones, and W.-S. Gan, ``Salsa: Spatial cue-augmented log-spectrogram features for polyphonic sound event localization and detection,'' \emph{IEEE/ACM Transactions on Audio, Speech, and Language Processing}, vol.~30, pp. 1749--1762, 2022.

\bibitem{piczak2015esc}
K.~J. Piczak, ``Esc: Dataset for environmental sound classification,'' in \emph{Proceedings of the 23rd ACM international conference on Multimedia}, 2015, pp. 1015--1018.

\bibitem{gemmeke2017audio}
J.~F. Gemmeke, D.~P. Ellis, D.~Freedman, A.~Jansen, W.~Lawrence, R.~C. Moore, M.~Plakal, and M.~Ritter, ``Audio set: An ontology and human-labeled dataset for audio events,'' in \emph{2017 IEEE international conference on acoustics, speech and signal processing (ICASSP)}.\hskip 1em plus 0.5em minus 0.4em\relax IEEE, 2017, pp. 776--780.

\bibitem{fonseca2021fsd50k}
E.~Fonseca, X.~Favory, J.~Pons, F.~Font, and X.~Serra, ``Fsd50k: an open dataset of human-labeled sound events,'' \emph{IEEE/ACM Transactions on Audio, Speech, and Language Processing}, vol.~30, pp. 829--852, 2021.

\bibitem{chen2020vggsound}
H.~Chen, W.~Xie, A.~Vedaldi, and A.~Zisserman, ``Vggsound: A large-scale audio-visual dataset,'' in \emph{ICASSP 2020-2020 IEEE International Conference on Acoustics, Speech and Signal Processing (ICASSP)}.\hskip 1em plus 0.5em minus 0.4em\relax IEEE, 2020, pp. 721--725.

\bibitem{fela2022perceptual}
R.~F. Fela, A.~Pastor, P.~Le~Callet, N.~Zacharov, T.~Vigier, and S.~Forchhammer, ``Perceptual evaluation on audio-visual dataset of 360 content,'' in \emph{2022 IEEE International Conference on Multimedia and Expo Workshops (ICMEW)}.\hskip 1em plus 0.5em minus 0.4em\relax IEEE, 2022, pp. 1--6.

\bibitem{dai2017scannet}
A.~Dai, A.~X. Chang, M.~Savva, M.~Halber, T.~Funkhouser, and M.~Nie{\ss}ner, ``Scannet: Richly-annotated 3d reconstructions of indoor scenes,'' in \emph{Proceedings of the IEEE conference on computer vision and pattern recognition}, 2017, pp. 5828--5839.

\bibitem{deitke2020robothor}
M.~Deitke, W.~Han, A.~Herrasti, A.~Kembhavi, E.~Kolve, R.~Mottaghi, J.~Salvador, D.~Schwenk, E.~VanderBilt, M.~Wallingford, L.~Weihs, M.~Yatskar, and A.~Farhadi, ``Robothor: An open simulation-to-real embodied ai platform,'' 2020.

\bibitem{huh2023epic}
J.~Huh, J.~Chalk, E.~Kazakos, D.~Damen, and A.~Zisserman, ``Epic-sounds: A large-scale dataset of actions that sound,'' in \emph{ICASSP 2023-2023 IEEE International Conference on Acoustics, Speech and Signal Processing (ICASSP)}.\hskip 1em plus 0.5em minus 0.4em\relax IEEE, 2023, pp. 1--5.

\bibitem{mohino2022introducing}
I.~Mohino-Herranz, J.~Garc{\'\i}a-G{\'o}mez, M.~Aguilar-Ortega, M.~Utrilla-Manso, R.~Gil-Pita, and M.~Rosa-Zurera, ``Introducing the realised dataset for sound event classification,'' \emph{Electronics}, vol.~11, no.~12, p. 1811, 2022.

\bibitem{chen2020soundspaces}
C.~Chen, U.~Jain, C.~Schissler, S.~V.~A. Gari, Z.~Al-Halah, V.~K. Ithapu, P.~Robinson, and K.~Grauman, ``Soundspaces: Audio-visual navigation in 3d environments,'' in \emph{Computer Vision--ECCV 2020: 16th European Conference, Glasgow, UK, August 23--28, 2020, Proceedings, Part VI 16}.\hskip 1em plus 0.5em minus 0.4em\relax Springer, 2020, pp. 17--36.

\bibitem{chen2022soundspaces}
C.~Chen, C.~Schissler, S.~Garg, P.~Kobernik, A.~Clegg, P.~Calamia, D.~Batra, P.~Robinson, and K.~Grauman, ``Soundspaces 2.0: A simulation platform for visual-acoustic learning,'' \emph{Advances in Neural Information Processing Systems}, vol.~35, pp. 8896--8911, 2022.

\bibitem{mei2023wavcaps}
X.~Mei, C.~Meng, H.~Liu, Q.~Kong, T.~Ko, C.~Zhao, M.~D. Plumbley, Y.~Zou, and W.~Wang, ``Wavcaps: A chatgpt-assisted weakly-labelled audio captioning dataset for audio-language multimodal research,'' \emph{arXiv preprint arXiv:2303.17395}, 2023.

\bibitem{Ephrat_2018}
\BIBentryALTinterwordspacing
A.~Ephrat, I.~Mosseri, O.~Lang, T.~Dekel, K.~Wilson, A.~Hassidim, W.~T. Freeman, and M.~Rubinstein, ``Looking to listen at the cocktail party: a speaker-independent audio-visual model for speech separation,'' \emph{ACM Transactions on Graphics}, vol.~37, no.~4, p. 1–11, July 2018. [Online]. Available: \url{http://dx.doi.org/10.1145/3197517.3201357}
\BIBentrySTDinterwordspacing

\bibitem{gao2019coseparating}
R.~Gao and K.~Grauman, ``Co-separating sounds of visual objects,'' 2019.

\bibitem{xu2019recursive}
X.~Xu, B.~Dai, and D.~Lin, ``Recursive visual sound separation using minus-plus net,'' 2019.

\bibitem{zhao2019sound}
H.~Zhao, C.~Gan, W.-C. Ma, and A.~Torralba, ``The sound of motions,'' 2019.

\bibitem{rouditchenko2019selfsupervised}
A.~Rouditchenko, H.~Zhao, C.~Gan, J.~McDermott, and A.~Torralba, ``Self-supervised audio-visual co-segmentation,'' 2019.

\bibitem{gan2020music}
C.~Gan, D.~Huang, H.~Zhao, J.~B. Tenenbaum, and A.~Torralba, ``Music gesture for visual sound separation,'' 2020.

\bibitem{oord2016wavenet}
A.~van~den Oord, S.~Dieleman, H.~Zen, K.~Simonyan, O.~Vinyals, A.~Graves, N.~Kalchbrenner, A.~Senior, and K.~Kavukcuoglu, ``Wavenet: A generative model for raw audio,'' \emph{Arxiv}, 2016.

\bibitem{richard2021neural}
A.~Richard, D.~Markovic, I.~D. Gebru, S.~Krenn, G.~A. Butler, F.~Torre, and Y.~Sheikh, ``Neural synthesis of binaural speech from mono audio,'' in \emph{International Conference on Learning Representations}, 2021.

\bibitem{donahue2019adversarial}
C.~Donahue, J.~McAuley, and M.~Puckette, ``Adversarial audio synthesis,'' in \emph{International Conference on Learning Representations}, 2019.

\bibitem{engel2019gansynth}
J.~Engel, K.~K. Agrawal, S.~Chen, I.~Gulrajani, C.~Donahue, and A.~Roberts, ``Gansynth: Adversarial neural audio synthesis,'' in \emph{International Conference on Learning Representations}, 2019.

\bibitem{chen2022visual}
C.~Chen, R.~Gao, P.~Calamia, and K.~Grauman, ``Visual acoustic matching,'' in \emph{Proceedings of the IEEE/CVF Conference on Computer Vision and Pattern Recognition}, 2022, pp. 18\,858--18\,868.

\bibitem{gao20192}
R.~Gao and K.~Grauman, ``2.5 d visual sound,'' in \emph{Proceedings of the IEEE Conference on Computer Vision and Pattern Recognition}, 2019, pp. 324--333.

\bibitem{zhou2020sepstereo}
H.~Zhou, X.~Xu, D.~Lin, X.~Wang, and Z.~Liu, ``Sep-stereo: Visually guided stereophonic audio generation by associating source separation,'' in \emph{Computer Vision--ECCV 2020: 16th European Conference, Glasgow, UK, August 23--28, 2020, Proceedings, Part XII 16}.\hskip 1em plus 0.5em minus 0.4em\relax Springer, 2020, pp. 52--69.

\bibitem{gao2019listen}
R.~Gao, T.-H. Oh, K.~Grauman, and L.~Torresani, ``Listen to look: Action recognition by previewing audio,'' \emph{arXiv preprint arXiv:1912.04487}, 2019.

\bibitem{tian2020unified}
Y.~Tian, D.~Li, and C.~Xu, ``Unified multisensory perception: Weakly-supervised audio-visual video parsing,'' in \emph{ECCV}, 2020.

\bibitem{zhou2020sep}
H.~Zhou, X.~Xu, D.~Lin, X.~Wang, and Z.~Liu, ``Sep-stereo: Visually guided stereophonic audio generation by associating source separation,'' in \emph{European Conference on Computer Vision}.\hskip 1em plus 0.5em minus 0.4em\relax Springer, 2020, pp. 52--69.

\bibitem{zhou2021pose}
H.~Zhou, Y.~Sun, W.~Wayne, C.~C. Loy, X.~Wang, and Z.~Liu, ``Pose-controllable talking face generation by implicitly modularized audio-visual representation,'' in \emph{Proceedings of the IEEE/CVF Conference on Computer Vision and Pattern Recognition}, 2021.

\bibitem{xu2021visually}
X.~Xu, H.~Zhou, Z.~Liu, B.~Dai, X.~Wang, and D.~Lin, ``Visually informed binaural audio generation without binaural audios,'' in \emph{Proceedings of the IEEE/CVF Conference on Computer Vision and Pattern Recognition}, 2021.

\bibitem{hu2020cross}
D.~Hu, X.~Li, L.~Mou, P.~Jin, D.~Chen, L.~Jing, X.~Zhu, and D.~Dou, ``Cross-task transfer for geotagged audiovisual aerial scene recognition,'' in \emph{ECCV}, 2020.

\bibitem{hu2020curriculum}
D.~Hu, F.~Nie, and X.~Li, ``Curriculum audiovisual learning,'' \emph{arXiv preprint arXiv:2001.09414}, 2020.

\bibitem{afouras2020self}
T.~Afouras, A.~Owens, J.~S. Chung, and A.~Zisserman, ``Self-supervised learning of audio-visual objects from video,'' in \emph{ECCV}, 2020.

\bibitem{qian2020multiple}
R.~Qian, D.~Hu, H.~Dinkel, M.~Wu, N.~Xu, and W.~Lin, ``Multiple sound sources localization from coarse to fine,'' in \emph{ECCV}, 2020.

\bibitem{hu2020discriminative}
D.~Hu, R.~Qian, M.~Jiang, X.~Tan, S.~Wen, E.~Ding, W.~Lin, and D.~Dou, ``Discriminative sounding objects localization via self-supervised audiovisual matching,'' in \emph{Advances in Neural Information Processing Systems}, 2020.

\bibitem{zhao2021_3DVG_Transformer}
L.~Zhao, D.~Cai, L.~Sheng, and D.~Xu, ``{3DVG-Transformer}: Relation modeling for visual grounding on point clouds,'' in \emph{ICCV}, 2021, pp. 2928--2937.

\bibitem{Luo_2022}
\BIBentryALTinterwordspacing
J.~Luo, J.~Fu, X.~Kong, C.~Gao, H.~Ren, H.~Shen, H.~Xia, and S.~Liu, ``3d-sps: Single-stage 3d visual grounding via referred point progressive selection,'' in \emph{2022 IEEE/CVF Conference on Computer Vision and Pattern Recognition (CVPR)}.\hskip 1em plus 0.5em minus 0.4em\relax IEEE, June 2022. [Online]. Available: \url{http://dx.doi.org/10.1109/CVPR52688.2022.01596}
\BIBentrySTDinterwordspacing

\bibitem{huang2022multi}
S.~Huang, Y.~Chen, J.~Jia, and L.~Wang, ``Multi-view transformer for 3d visual grounding,'' in \emph{Proceedings of the IEEE/CVF Conference on Computer Vision and Pattern Recognition}, 2022, pp. 15\,524--15\,533.

\bibitem{wu2022wav2clip}
H.-H. Wu, P.~Seetharaman, K.~Kumar, and J.~P. Bello, ``Wav2clip: Learning robust audio representations from clip,'' in \emph{ICASSP 2022 - 2022 IEEE International Conference on Acoustics, Speech and Signal Processing (ICASSP)}, 2022.

\bibitem{qi2019deep}
C.~R. Qi, O.~Litany, K.~He, and L.~J. Guibas, ``Deep hough voting for 3d object detection in point clouds,'' in \emph{proceedings of the IEEE/CVF International Conference on Computer Vision}, 2019, pp. 9277--9286.

\bibitem{touvron2023llama}
H.~Touvron, T.~Lavril, G.~Izacard, X.~Martinet, M.-A. Lachaux, T.~Lacroix, B.~Rozi{\`e}re, N.~Goyal, E.~Hambro, F.~Azhar, \emph{et~al.}, ``Llama: Open and efficient foundation language models,'' \emph{arXiv preprint arXiv:2302.13971}, 2023.

\bibitem{radford2021learning}
A.~Radford, J.~W. Kim, C.~Hallacy, A.~Ramesh, G.~Goh, S.~Agarwal, G.~Sastry, A.~Askell, P.~Mishkin, J.~Clark, G.~Krueger, and I.~Sutskever, ``Learning transferable visual models from natural language supervision,'' 2021.

\bibitem{gadre2022cow}
S.~Y. Gadre, M.~Wortsman, G.~Ilharco, L.~Schmidt, and S.~Song, ``Cows on pasture: Baselines and benchmarks for language-driven zero-shot object navigation,'' \emph{CVPR}, 2023.

\bibitem{Chefer_2021_CVPR}
H.~Chefer, S.~Gur, and L.~Wolf, ``Transformer interpretability beyond attention visualization,'' in \emph{Proceedings of the IEEE/CVF Conference on Computer Vision and Pattern Recognition (CVPR)}, June 2021, pp. 782--791.

\bibitem{kamath2021mdetr}
A.~Kamath, M.~Singh, Y.~LeCun, G.~Synnaeve, I.~Misra, and N.~Carion, ``Mdetr -- modulated detection for end-to-end multi-modal understanding,'' 2021.

\bibitem{carion2020endtoend}
N.~Carion, F.~Massa, G.~Synnaeve, N.~Usunier, A.~Kirillov, and S.~Zagoruyko, ``End-to-end object detection with transformers,'' 2020.

\bibitem{minderer2022simple}
M.~Minderer, A.~Gritsenko, A.~Stone, M.~Neumann, D.~Weissenborn, A.~Dosovitskiy, A.~Mahendran, A.~Arnab, M.~Dehghani, Z.~Shen, X.~Wang, X.~Zhai, T.~Kipf, and N.~Houlsby, ``Simple open-vocabulary object detection with vision transformers,'' 2022.

\bibitem{zhou2023esc}
K.~Zhou, K.~Zheng, C.~Pryor, Y.~Shen, H.~Jin, L.~Getoor, and X.~E. Wang, ``Esc: Exploration with soft commonsense constraints for zero-shot object navigation,'' 2023.

\end{thebibliography}
}
\addtolength{\textheight}{-12cm}   





\end{document}